\documentstyle[12pt,psfig]{article}
\newcommand{\be}{\begin{equation}}
\newcommand{\ee}{\end{equation}}
\newcommand{\bea}{\begin{eqnarray}}
\newcommand{\eea}{\end{eqnarray}}
\newcommand{\Hdip}{{\cal H}_{\hbox{\scriptsize dip}}}
\newcommand{\Han}{{\cal H}_{\hbox{\scriptsize an}}}
\newcommand{\Hex}{{\cal H}_{\hbox{\scriptsize ex}}}
\begin{document}
\title{Domain structures in ultrathin magnetic films}
\author{Paolo Politi}
\maketitle
\begin{center}
Dipartimento di Fisica dell'Universit\`a degli Studi di Firenze and Istituto 
Nazionale di Fisica della Materia, L.go E. Fermi 2, I-50125 Firenze, Italy\\
E-mail: politip@fi.infn.it
\end{center}
%\date{today}
\begin{abstract}
Domain structures appear in ultrathin magnetic films, 
if an easy-axis anisotropy, favouring the direction perpendicular to the
film, overcomes the shape anisotropy.
We will study the nature of the magnetic ground state and how the different
parameters influence the domain structure.
Both Ising and Heisenberg spins will be considered: in the latter case,
a reorientation phase transition may take place. 
Domains are expected to appear just before such transition.
Domain structures at finite temperature will be described in terms of a 
two dimensional array of domain walls, whose behaviour resembles a
liquid crystal.
\end{abstract}

\section{Introduction}
A three dimensional ferromagnetic sample, below its Curie temperature, 
generally has no net magnetization because $-$as suggested by Weiss~%
\cite{Weiss} and discovered by Bitter~\cite{Bitter}$-$ it is made up of
contiguous regions where the magnetization is oriented along different
directions. These regions (typical size $\approx 1\div 10\mu$m) are
called magnetic domains, and are separated by domain walls (typical size
$\approx 10^2\div 10^3$\AA), where the magnetization rotates
continuously from one direction to another.

The magnetization inside each domain is due to the ferromagnetic coupling
$(\Hex)$ between neighboring spins, which aligns them parallely. The
existence of domains is due to the dipolar energy $(\Hdip)$: the true
``magnetic" (and long-range) interaction between spins. This energy is also
called ``magnetostatic energy" or ``shape anisotropy": in fact, its
minimization strongly depends on the shape of the sample.
Finally, domains are not magnetized in random directions, but they prefer
certain directions, depending on the crystal-axes, whose existence is due
to the so-called magnetocrystalline anisotropies $(\Han)$.
The actual magnetic structure of a system depends on the minimization
of the total energy ${\cal H} = \Hex + \Hdip + \Han$.

In this paper, we will be interested in ultrathin magnetic films:
by ``ultrathin" we mean a thickness smaller than, or of order of 
$N=10$ atomic monolayers. Such small thicknesses can be currently
obtained through growth by Molecular Beam Epitaxy~\cite{APA}.
Our aim is to give a microscopic description of domain structures
in such two-dimensional systems, both at zero temperature
(Section~\ref{T=0}) and at finite temperature (Section~\ref{finiteT}).
Section~\ref{LvsN} is devoted to the dependence of the
domain size $L$ on the film thickness $N$. A comparison with 
Kittel's theory $-$valid for much thicker films ($N>L$)$-$ will be done.

Ultrathin films can undergo a reorientation transition, with the
magnetization passing from being perpendicular to the film to being parallel
to it. Since the condition for the occurence of this transition
coincides with the condition for having domains not unphysically large, the
interplay between reorientation transition and magnetic domains will
be treated in Section~\ref{RPT}.

\section{Domain structures at zero temperature}
\label{T=0}

The aim of the present section will be the study of the classical
ground state of a ferromagnetic ultrathin film, in presence of dipolar
interactions and of a uniaxial anisotropy. Here and in the
following, both the spin modulus $|\vec S|$ and the lattice constant $a$ 
of the cubic crystal will be put equal to one. 

The hamiltonian of a single monolayer will generally be assumed to have the form

\begin{equation}
{\cal H} = \Hex + \Han + \Hdip
\label{2.1} 
\end{equation}
where the first term
\begin{equation}
\Hex   = {J\over 2} \int d\vec x\left(\nabla\vec S\right)^2\nonumber
\label{2.2} 
\end{equation}
is the Heisenberg interaction, in the continuum approximation. The second term
\begin{equation}
\Han     = -{K\over 2}\int d\vec x S_z^2(\vec x)\nonumber\\ 
\label{2.3} 
\end{equation}
is a  single-ion
anisotropy caused by the crystal field; it appears because, in 
a thin film, the $\hat z$ direction (perpendicular to the film) is not
equivalent to the in-plane directions. The constant $K$ may be
negative (easy-plane effect), but in this case 
no domain structure appears. Therefore, only the opposite case $(K>0)$
will be considered. Finally, the third term
\begin{equation}
\Hdip    =\Hdip^{(1)}+\Hdip^{(2)}
\label{2.4} 
\end{equation}
is the dipolar interaction, which can conveniently be split into a 
rotationally invariant part
\begin{equation}
\Hdip^{(1)}  = {\Omega\over 8\pi}
\int {d\vec x d\vec x'\over |\vec x-\vec x'|^3}
 \vec S(\vec x)\cdot\vec S(\vec x')
\label{2.5} 
\end{equation}
and an anisotropic part
\begin{equation}
\Hdip^{(2)}  = -{3\Omega\over 8\pi}
\int {d\vec x d\vec x'\over |\vec x-\vec x'|^3}
(\vec S(\vec x)\cdot\vec\nu)(\vec S(\vec x')\cdot\vec\nu)
\label{2.6} 
\end{equation}
which couples the spin space to the ``real" space. 
In addition to the constant $\Omega$, we have introduced the notation:
$\vec\nu\equiv (\vec x-\vec x')/|\vec x-\vec x'|$.

Since a spin doesn't interact with itself, the integration must exclude
a region around $x=x'$. This problem doesn't exist if a discrete calculation
replaces the integration (see below), or in the calculations
relevant for domain structures, where $x$ and $x'$ are always far away (because
belonging to different domains).

The term $\Hdip^{(1)}$ represents a long-range,
antiferromagnetic coupling, while $\Hdip^{(2)}$
introduces an  easy-plane anisotropy. In fact,
this term vanishes for a perpendicular collinear state (hereafter called
$\perp$ state), because $\vec S\cdot\vec\nu\equiv 0$, whilst for
a parallel collinear state (hereafter called $\parallel$ state), 
it has the negative value:
\be
-{3\Omega\over 8\pi}\sum_{\vec x}{(\vec S\cdot\vec\nu)^2\over|\vec x|^3}
\label{dipani}
\ee
per spin. So, the
$\parallel$ configuration is energetically favoured with respect to the
$\perp$ one; furthermore, since only the in-plane component $\vec S_\parallel$
of the spin contributes to it, for a generic collinear configuration
forming an angle $\theta$ with the $\hat z$ axis, 
$(\vec S\cdot\vec\nu)^2$ rewrites as  $(\vec S_\parallel\cdot\vec\nu)^2$ and
therefore expression~(\ref{dipani}) 
simply needs to be multiplied by $\sin^2\theta$.
So, for a collinear configuration, $\Hdip$ is found to reduce $-$up to a
constant$-$ to the anisotropic expression:
\be
\Hdip^{an} = -{\Omega\over 2}\sin^2\theta\int d\vec x =
{\Omega\over 2}\int d\vec x S_z^2
\label{H_dip^an}
\ee

Sometimes, instead of $\Omega$ we will use the quantity
$g\equiv\Omega/16\pi$, which appears in the discrete version of 
dipolar interaction:
$$
\Hdip=g\sum_{i>j}{1\over r_{ij}^3}
\left[\vec S_i\cdot\vec S_j -3(\vec S_i\cdot\nu)
(\vec S_j\cdot\nu)\right]~.
$$

Because of the competition between $\Hdip^{(1)}$ (a weak long-range
antiferromagnetic (AFM) coupling) and $\Hex$ (a strong short-range
ferromagnetic (FM) coupling), a domain structure
arises, for {\it any} value of the dipolar coupling;
furthermore, the competition between $\Hdip^{(2)}$ 
(an effective easy-plane anisotropy) 
and $\Han$ (an easy-axis anisotropy) can produce a 
reorientation phase transition (RPT) of the magnetization, from 
the $\perp$ state to the $\parallel$ state. 
To prove the former statement, it is sufficient
to show that the $\perp$ state is destabilized by the introduction
of an Ising domain wall. 
So, if we invert in the negative $\hat z$ direction all the
spins in the half-plane $x>0$, we have to compare the cost of the domain
wall ($2J$ per unit length) with the energy gain, due to dipolar
interaction. Such gain is obtained by summing up $(-2g/|\vec x-\vec x'|^3)$
on all the spins $(\vec x')$ of the half-plane $x>0$ and the spins $(\vec x)$ 
of the half-plane $x<0$, with $y=$~constant (because we need the gain 
per unit length). The first sum on $\vec x'$
corresponds to two spatial integrations, giving a quantity of order $-2g/|x|$.
Therefore, the third sum diverges logarithmically at large distances%
\footnote{At small distances, the lattice constant is the natural cut-off.}
whatever the value of $g$, thus proving that the
dipolar gain always prevails on the exchange loss.

The previous rough calculation also allows to have a semi-quantitative
expression for the domain size of a stripe domain structure. 
In fact, the third
integration on $x$ will be limited up to the domain size $L$, by giving
an energy gain of order $-g\ln L$ which must be approximately equal to
the domain wall energy (per unit length): $g\ln L\approx E_{dw}$. So:
\be
L\approx\exp(E_{dw}/g)~.
\ee

For an Ising system, $E_{dw}=2J$, so that (at zero temperature: $T=0$) 
the domain size is exponentially large, unless $J$ and $g$ 
are of the same order. In Section~\ref{finiteT}, we will see how the previous
result is modified, if thermal effects are taken into account.

For an Heisenberg system, the domain wall energy has the form~\cite{LKittel}:
$E_{dw}=2\sqrt{\lambda J}$, where $\lambda$ is the
{\it total} effective easy-axis anisotropy (see Eq.~(\ref{H_dip^an})): 
$\lambda=K-\Omega$. Therefore, domains of ``small" extension 
can be obtained even if $J\gg K,\Omega$, provided that $K\simeq\Omega$.

Our calculations will concern a square lattice. A triangular lattice gives rise
to frustration, in presence of an antiferromagnetic exchange coupling.
Indeed, dipolar coupling is an antiferromagnetic interaction, in the Ising
limit; anyway, it is a long-range one: so, the induced frustration exists even 
in a square lattice. We expect that the picture discussed in the rest of this 
Section is not qualitatively modified for a triangular lattice,
if $L\gg 1$.

\subsection{Domain walls}
\label{ss_dw}
The size of a domain wall is given by 
the balance between the effective easy-axis
anisotropy, and the exchange energy. If the long-range character of
$\Hdip$ is neglected and only its easy-plane effect (\ref{H_dip^an}) 
is retained, it is possible to determine analytically the 
magnetization profile inside the wall, and the domain wall extension.

If the wall is parallel to the $y$ axis (as it will be always
supposed in the following), in the continuum approximation  the domain
wall energy (per unit length) is
\be
{\cal H}_{dw} = {1\over 2}\int dx \left[ 
J(\nabla\vec S)^2 -\lambda S_z^2\right] ~,
\label{H_dw}
\ee
which must be minimized with respect to $\vec S(x)$. 

By using polar
coordinates $(\vec S=(\sin\theta\cos\phi,\sin\theta\sin\phi,\cos\theta))$
we obtain
\be
{\cal H}_{dw} = \int {dx\over 2} \left\{ 
J[(\partial_x\theta)^2 +(\partial_x\phi)^2\sin^2\theta]
-\lambda\cos^2\theta\right\}~.
\ee
The $\phi$-dependent term is minimized when $\phi$ is constant,
e.g. $\phi=\pi/2$ (Bloch wall) or $\phi=0$ (N\'eel wall).
On the other side, the minimization with respect to $\theta$ gives
$2\partial^2_{xx}\theta =(\lambda/J)\sin(2\theta)$, whose solution gives
$\theta(x)=2\arctan\left(e^{x/w}\right)$, where $w=\sqrt{J/\lambda}$.
The corresponding expression of $E_{dw}$ is obtained from Eq.~(\ref{H_dw}):
$E_{dw}=J/w +w\lambda=2\sqrt{\lambda J}$.

The degeneracy among different types of wall is removed if dipolar
interaction is correctly taken into account.
It is immediately seen that the relevant interaction is the self-energy
of the domain wall: in fact, let us consider the dipolar interaction between
a spin $\vec S_1$ belonging to the wall, and a spin $\vec S_2$ belonging
to a nieghbouring domain. Since $\vec S_2\cdot\vec\nu=0$ ($\vec\nu=
\vec r/r$, $\vec r$ being the vector joining the two spins)
only the isotropic interaction $g(\vec S_1\cdot\vec S_2)/r^3$ is not 
vanishing. Anyway, the quantity $\vec S_1\cdot\vec S_2=\cos\theta_1$
does not discriminate among different types of wall, because it only depends 
on the axial angle $\theta_1$ of the spin $\vec S_1$, whilst walls
differ in the polar angle $\phi_1$.

The dipolar contribution to the domain wall energy therefore writes:
\be
E_{dw}^{dip}={g\over 2}\sum_{x=0}^w \sum_{x'=0}^w \sum_{y=-\infty}^\infty 
\left\{ {\vec S(x)\cdot\vec S(x')\over [(x-x')^2 +y^2]^{3/2} }
-3 { (\vec S(x)\cdot \vec r)(\vec S(x')\cdot\vec r)\over
[(x-x')^2 +y^2]^{5/2} } \right\}~,
\ee
where we have used the translational invariance in the $\hat y$ direction, and
the following notations are understood: $\vec r=(x-x',y)$ and
$\vec S(x)=(\sin\theta(x)\cos\phi,\sin\theta(x)\sin\phi,\cos\theta(x))$.
The function $\theta(x)$ is any increasing monotonic function, such that
$\theta(0)=0$ and $\theta(w)=\pi$. Our aim is to minimize
$E_{dw}^{dip}$ with respect to the polar angle $\phi$.

It is straightforward to obtain the following relations:
\bea
\vec S(x)\cdot\vec S(x') &=& \cos(\theta(x)-\theta(x'))\nonumber\\
(\vec S(x)\cdot\vec r)(\vec S(x')\cdot\vec r) &=& \sin\theta(x)\sin\theta(x')
[(x-x')^2\cos^2\phi +y^2\sin^2\phi +2y(x-x')\sin\phi\cos\phi]\nonumber
\eea

By retaining only the $\phi$-dependent terms, we have:
\bea
E_{dw}^{dip}=-{3g\over 2}\sum_{x=0}^w \sum_{x'=0}^w \sum_{y=-\infty}^\infty  &&
{\sin\theta(x)\sin\theta(x')\over [(x-x')^2 +y^2]^{5/2} }\times\nonumber\\
&&\left\{ \sin^2\phi [y^2-(x-x')^2] +y(x-x')\sin 2\phi \right\}~.\nonumber
\eea

The term proportional to $\sin 2\phi$ gives a zero contribution, because it
is an odd function of $y$. So, we are left with the expression:
\bea
E_{dw}^{dip}(\phi)&=&-{3g\over 2}\sin^2\phi
\sum_{x=0}^w \sum_{x'=0}^w \sum_{y=-\infty}^\infty 
\sin\theta(x)\sin\theta(x') {y^2-(x-x')^2\over [(x-x')^2 +y^2]^{5/2} }
\nonumber\\
&\equiv& -{3g\over 2}\sin^2\phi\cdot {\cal I}(w)\nonumber
\eea

Generally speaking, the shape anisotropy due to dipolar interaction
implies that two spins $\vec S(\vec x)$, $\vec S(\vec x')$ will prefer
to align parallely to $(\vec x-\vec x')$. Therefore
the sum ${\cal I}(w)$ simply reflects the
shape anisotropy of a stripe: $\overline{y^2}>\overline{(x-x')^2}$,
where the bar means some average on the stripe. 
In other words, the dipolar energy of the domain wall is minimized
if the spins point on the average in the (elongate) $\hat y$ direction.
The positive factor $\sin\theta(x)\sin\theta(x')$ does not play a 
major role: it is present because the in-plane component of the spin
varies inside the wall.
In conclusion, ${\cal I}(w)$ is positive and
$E_{dw}^{dip}$ is minimal for $\phi=\pi/2$, i.e. 
for a  Bloch-type wall.

\subsection{The Ising model}
Once we have shown that the ferromagnetic configuration is 
destabilized by dipolar
interactions, whatever the value of $g$, in this section we will
study more closely the Ising model, in the entire range of values for the
ratio $g/J$. Let us start with the physically relevant case
$g/J\ll 1$, which corresponds to very large domains.

By extending a little more the na\"\i ve calculation made in the introduction,
it is possible to determine the {\it leading} term in the expression of the
dipolar energy of a stripe configuration. 
In order to use the following results also for the Heisenberg model, and
for a larger number of atomic planes, we will suppose the two neighbouring
stripes of size $L$ to be separated by a domain wall of size $w\ll L$ and to
lie on different planes at distance $c$. So, the relevant integral is the
following:
\be
I(c)=- \int_0^{L-w} dx' \int_0^L dx \int_{-\infty}^{+\infty} dy
{1\over\left[ (x+x')^2 + y^2 +c^2 \right]^{3/2} }~,
\label{Ic_1}
\ee
whose general solution is:
\bea
I(c) &=& \left({4L-2w\over c}\right)\arctan\left({2L-w\over c}\right)
-\ln\left[ 1+\left({2L-w\over c}\right)^2\right]  
-\left({4L\over c}\right)\arctan\left({L\over c}\right) \nonumber\\
&& +2\ln\left(1+{L^2\over c^2}\right)
+{2w\over c}\arctan\left({w\over c}\right)
-\ln\left(1+{w^2\over c^2}\right)~.
\label{Ic_2}
\eea

In the case $c=0$ (corresponding to a single monolayer) 
or for $w\gg c$, we find $I(c)\approx 2\ln(L/2w)$. 
In these cases, the dipolar energy gain with respect to 
the $\perp$ state is, for a couple of planes at distance $c$:
\be
E_{dip}^{2~planes} = -2g I(c) = -4g\ln\left({L\over 2w}\right)~.
\ee
For an Ising monolayer, $c=0$ and $w=1$, the total energy per spin
of a stripe (S) configuration
with respect to the $\perp$ state is: 
\be
E^S(L) = {2J\over L} - {4g\over L}\ln(L/2)~.
\ee
However, the factor 2 inside the logarithm 
(which modifies, in a certain sense, the value of $J$: $J\rightarrow J+
2g\ln 2$) is not meaningful, because it contributes to the non leading
term of the dipolar energy. Conversely, it is noteworthy that the leading term 
(proportional to $g\ln L$) is the one obtained by considering the interaction
(per unit length) of a stripe of size $L$ with a neighboring 
stripe of infinite size!
Analogously, the same leading term appears if the neighbouring 
stripe is indeed a
square of size $L$. This result therefore provides $-$in the same
approximation$-$ the expression $E^C(L)$
for the energy of a square network (C=chess-board) of size $L$:
\be
E^C(L) = {4J\over L} - {8g\over L}\ln(L) \equiv 2 E^S(L)~.
\ee
Therefore, not only the exchange cost of a square network is the double of
a stripe structure with the same period
(as the total length of domain walls is the double),
but also the dipolar gain is the double. 
In very poor terms, the reason is that the infinite size of a stripe
contributes only to the next-to-leading order term. Since in a square
network each domain has four neighboring domains with opposite magnetization,
whilst in a stripe configuration only two, we have the result
$E^C(L)=2E^S(L)$ (at the leading order in the dipolar energy!).

The previous results have been found by Czech and Villain~\cite{CV} (see
their Eq.~(12)), who have drawn the erroneous conclusion that the chess-board
configuration has a lower energy than the stripe configuration.
Indeed, the next-to-leading order terms modify the previous picture,
as explained by Kaplan and Gehring~\cite{KG}.
Through a more careful evaluation of the dipolar sums, they find that
$E^S(L)$ and $E^C(L)$ are modified as follows:
\bea
E^S(L) &=& {2J-\alpha_S g\over L} - {4g\over L}\ln L \nonumber\\
E^C(L) &=& {4J-\alpha_C g\over L} - {8g\over L}\ln L \nonumber
\eea
where $\alpha_S\simeq 4.2$ and $\alpha_C\simeq 0.3$. By minimization with
respect to $L$, we find
\bea
&& L_S = \exp\left[{J\over 2g} +1 -{\alpha_S\over 4}\right]~~~~~~
E^S(L_S) = - {4g\over L_S} \nonumber\\
&& L_C = \exp\left[{J\over 2g} +1 -{\alpha_C\over 8}\right]~~~~~~
E^C(L_C) = - {8g\over L_C} \nonumber
\eea

From the previous expressions, it is apparent why neglecting terms of order
$(g/L)$ in the energy leads to mistakes. Furthermore, $E^S<E^C$ if
$L_C/L_S>2$, that is to say $(\alpha_S/4-\alpha_C/8)>\ln 2$, a relation which 
is satisfied.

Now, we will consider the opposite limit of a large ratio $g/J$:
dipolar interaction prevails over the FM exchange.
In the following, the notations C-$L$ and S-$L$ will mean, respectively a
square network of side $L$ and a stripe configuration of period $L$.
Since dipolar interaction is a (long-range) AFM coupling, we are led to
conclude that the AFM state (i.e. the C-1 state) is the lowest
energy one: an explicit numerical calculation~\cite{MacIsaac} shows
that the C-1 state is indeed the ground state.

Such a state is maintained for a sufficiently weak $J$. So, a
square network is preferred at (very) high $g/J$, but stripe 
configurations are preferred at low $g/J$.
We can ask when the transition between square configurations and stripe ones
takes place. According to numerical results~\cite{MacIsaac}, for
$J\simeq 0.4g$, the C-1 state is replaced by the S-1 state, and for
increasing $J$ we simply have an augmentation of the period $L$ of the
stripe configuration. Since analytical 
theories are not applicable here, let us give some
simple arguments to justify why C-1 is replaced by S-1, rather than C-2.
Let us compare the energies of the two possible configurations:
\bea
E^C(2) &=& E_{dip}^C(2) + 4J/2\nonumber\\
E^S(1) &=& E_{dip}^S(1) +2J~.\nonumber
\eea
So, it is sufficient to compare their dipolar energies. 
Both the configurations have a vanishing nearest bond
contribution to the dipolar energy (two spins up and two spins down), but
in the S-1 state all the four nn-bonds are AFM, 
whilst in C-2 are half and half.
For the 3$^{rd}$ nearest bonds, the picture is reversed, but the coupling is 
weaker, and up to the 6$^{th}$ nearest bonds the contributions are the same.
This simple evaluation gives the correct result.

We conclude this section by summing up the main results.
Dipolar interactions favour an AFM coupling
between spins; if we consider chess-board (C-$L$) and stripes (S-$L$)
configurations, we have that the gains of dipolar energy (with respect
to the $\perp$ state) $|E_{dip}^S(L)|$ and $|E_{dip}^C(L)|$ are 
decreasing
functions of $L$ and $|E_{dip}^C(L)|>|E_{dip}^S(L)|$. Nevertheless, when a
FM exchange interaction is taken into account, the stripe configuration is 
energetically favoured, except for $J/g<0.4$, when the ground state is the
AFM one.

\subsection{The Heisenberg model}
In the previous section we have supposed that spins are ``Ising spins":
$\vec S$ is always parallel to the $\hat z$ axis. Now spins are allowed
to rotate, and the main difference is the energy of
the domain wall, which is lowered with respect to the Ising case. This
allows smallest domain sizes, even if $-$at $T=0-$ 
they will be seen to remain of
macroscopic extent, except in a very narrow region of the parameter
space. A second important feature of the ferromagnetic
Heisenberg model, is that its excitations are spin-waves, which allow
a local linear stability analysis. 

Some insight in the model is given by the study of the
collinear configurations, the generic one forming an angle $\theta$ with
the $\hat z$ axis. Since the total effective uniaxial
anisotropy is $\lambda=(K-\Omega)$, the $\perp$ state ($\theta=0$) 
will be the lowest energy one for $K>\Omega$. For $K<\Omega$ the 
magnetization will lie in the plane ($\theta=\pi/2$), and for
$K=\Omega$ all the collinear states will be degenerate.

The previous section has been devoted to show (for an Ising model) how 
the long-range character of the dipolar interaction ``destabilizes"
the $\perp$ state. This result is certainly not modified for an
Heisenberg model, which simply allows for a much larger phase space.
Anyway, there is no reason
to think that an infinitesimal variation in the $\perp$ state will lead to a
lower energy: in other words, the $\perp$ state may be locally stable and
spin-waves do not destabilize it, i.e. it may be a metastable state. 
In fact, the dispersion curve writes:
\be
\omega_\perp(q)=(K-\Omega)+Jq^2=\lambda +Jq^2~.
\ee
Dipolar interaction simply modifies the gap, because of its ``easy-plane"
effect.

If now we turn to the $\parallel$ state $(K<\Omega)$, a first question is:
Do dipolar interactions destabilize the $\parallel$ state? The
answer is no: this case, domains are not energetically favoured. 
In the $\perp$ state, domains rise because of the competition between the
FM $\Hex$ and the AFM $\Hdip^{(1)}$. Anyway, in the $\parallel$ state, the
anisotropic term $\Hdip^{(2)}$ (which vanishes almost everywhere in the
$\perp$ state) changes the character of $\Hdip$. Let us compare the
in-plane FM and AFM configurations. 

The spin in the origin is choosen
along the $\hat y$ axis: this way, all the other spins can be arranged
in circular rings of increasing radium where spins are mutually parallel,
and $-$in the FM state$-$ always oriented along the $\hat y$ direction, 
or $-$in the AFM state$-$ alternately oriented along  $\hat y$ and $-\hat y$.
An in-plane domain structure may arise only if $\Hdip$ favours the AFM
configuration with respect to the FM one. To check it, let us consider
the dipolar interaction between the spin $i$ in the origin and all
the spins $j$ on a given circular ring at distance $r$.
The resulting sum is proportional to 
$\pm[1-3\langle(\vec S\cdot\vec\nu_{ij})^2\rangle_j]/r^3$, where the sign
$+$ (resp.~$-$) holds for the FM (resp.~AFM) configuration. Since the average
$\langle\dots\rangle_j$ on the spins $j$ gives $(1/2)$, 
the negative anisotropic part prevails on the first one, and 
the energy is lower for the FM state! Consequently, domains are not
expected if the magnetization lies in the film plane. 

Anyway, the linear stability analysis of the $\parallel$
state reveals important features. In fact, 
it was shown by Maleev~\cite{Maleev} in the seventies that dipolar
interactions are able to stabilize the $\parallel$ 
state at finite temperature,
because of their long-range character. This means that the dispersion curve
is modified well beyond the introduction of some effective easy-plane
anisotropy, which is unable to assure a finite magnetization.
The correct expression for the dispersion curve is~\cite{Maleev,Bruno,PPR}
($\theta_{\vec q}$ is the in-plane angle between $\vec q$ and the
magnetization):
\be
\omega^2_\parallel(q)=\left[ (\Omega-K) -{\Omega\over 2}q + Jq^2\right]\times
\left[ {\Omega\over 2}q\cos^2\theta_{\vec q} +Jq^2\right]
\equiv \omega_{out}\times\omega_{in}~.
\label{w_par}
\ee
We have assigned a different meaning to the two terms, because now
the Goldstone mode $(\omega(q=0)=0)$ exists for 
$\omega_{in}$, but not for $\omega_{out}$. Anyway, an out of plane excitation
with a {\it finite} wavevector may go to zero and destabilize the
$\parallel$ configuration. This happens for $\bar q=\Omega/4J$, when
$1>(K/\Omega)>1-(\Omega/16J)$. So, for a weak dipolar 
interaction such interval is extremely narrow.

We can summarize the results of the spin-wave stability analysis of the FM
configurations, as follows. The $\perp$ state is locally stable, and
therefore it is metastable, because the insertion of a domain wall
lowers the energy of the system. Domains are not expected in the
in-plane state; nonetheless, the long-range character of $\Hdip$
destabilizes the $\parallel$ configuration, if $\Omega$ is just a bit larger
than $K$.
The main result of the next section will be to link quantitatively 
the results of the previous analysis, with energetic calculations on
domain structures.

\subsection{Yafet and Gyorgy's theory}
To my knowledge, the only microscopic study of domain strutures in a
Heisenberg system, is due to Yafet and Gyorgy~\cite{YG}. They study a
stripe domain structure, characterized by two parameters: the domain
size $L$ and the domain wall size $w$. The wall is of Bloch-type, and
the component $S_z$ of the magnetization,
along the domain wall, has a sinusoidal profile.
As suggested by the same authors, we have added a third parameter which reveals
essential to match the results of the 
spin-wave stability analysis:
the maximal component $s$ of the magnetization in the $\hat z$ direction, inside
each domain. $s=0$ means the $\parallel$ configuration, while $s=1$
is a ``standard" domain structure.
The energy of a generic stripe domain structure is therefore a function of the
three variational parameters $(L,w,s)$. Its minimization, and a comparison
with the energy of a generic collinear state, will allow to determine
the lowest energy configuration.

\begin{figure}
\centerline{\psfig{%
bbllx=60bp,bblly=100bp,bburx=540bp,bbury=680bp,figure=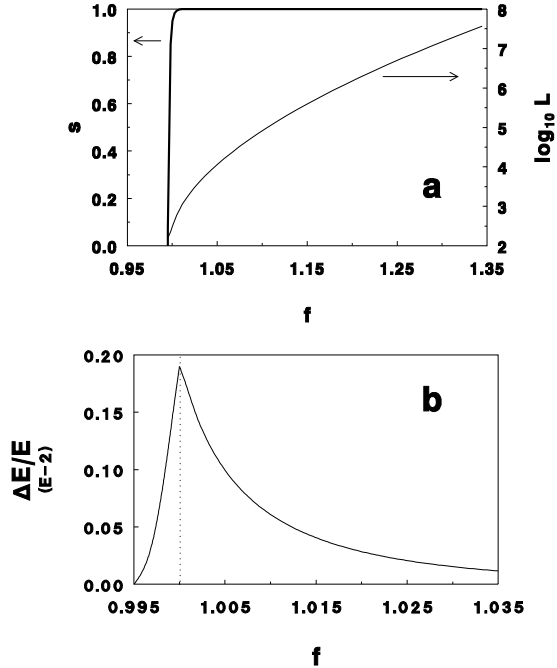,%
height=10cm,angle=0}}
\caption{(a) The maximal value $s$ of the component $|S_z|$, inside each
domain, and the domain size $L$ (in logarithmic scale).
(b)~~The energy gain of the domain structure, relative to the collinear state
(the $\perp$ one for $f>1$ and the $\parallel$ one for $f<1$).
The smallest value of $f$ shown in (b) is $f_m$, to which corresponds
$-$in (a)$-$ the vanishing of $s$.
For both figures,
$2L_d=25$. Similar dependence for $L(f)$ and (b) had already been given in
Ref.~\protect\cite{YG}.}
\label{fig_YG}
\end{figure}

In Fig.~\ref{fig_YG} we plot (a) $L$ and $s$, 
as a function of the ratio $f\equiv K/\Omega$, and (b)
the energy gain of the domain structure. 
Two features are manifest: (i)~It exists a minimal value $f_{m}<1$, beyond
which a domain structure is energetically favoured. 
As expected, the energy gain
persists till $f=\infty$ and it is maximal for $f=1$. 
(ii)~When $f=f_m$, spins lie in the plane ($s=0$), and $L=L_m=4\pi L_d$,
where $L_d=J/\Omega$ is the dipolar length. Upon
increasing $f$, the maximal component of $S_z$ immediately saturates
to $s=1$, and the domain size $L$ grows exponentially: unless dipolar
coupling and uniaxial anisotropy nearly compensates, domains are
unphysically large. 
How close to one $f$ must be, in order to observe domains,
it depends on the dipolar length.
It is not possible to give here the details of the calculation, which
follow Ref.~\cite{YG}. Anyway, the introduction of the parameter $s$
modifies the values of $f_m$ and $L_m$ found by Yafet and Gyorgy, which now
match the corresponding values, determined {\it via} the linear stability
analysis of the $\parallel$ configuration~\cite{Tesi,Libro}:
$f_m=1-(\Omega/16J)$ and $L_m=\pi/\bar q$ (see below Eq.~(\ref{w_par})).

In conclusion, the following picture emerges:
If $K$ is fairly smaller than $\Omega$, 
the $\parallel$ configuration is the ground
state; when $K$ is nearly equal to $\Omega$ ($K=\Omega-\Omega^2/16J$) 
a soft mode appears,
with a finite wavevector $\bar q=\pi/L_m$. In the new configuration, spins
acquire a component $S_z$ which is alternately positive and negative.
Upon further increase of $K$, a domain structure clearly appears, with
regions where $S_z=\pm 1$ are separated by domain walls of size $w$.
These configurations are the true ground state, {\it but} for $K>\Omega$ the
$\perp$ state is locally stable. The domain size $L$ increases
exponentially and rapidly becomes larger than the sample size.

The main difference between Ising and Heisenberg models is the form, and
therefore the cost, of a domain wall: in the former case, the wall
size is one lattice constant, and its energy per unit length is
simply $E^{Ising}_{dw}=2J$. 
In the latter case, the wall can extend over many lattice
constants and its cost is given by the balance between easy-axis anisotropy
and exchange: $E^{Heisenberg}_{dw}=2\sqrt{J(K-\Omega)}$. 
This expression correctly shows that 
the domain wall energy goes to zero, when $K$ and $\Omega$ nearly compensate.
Nevertheless, it has the drawback not to match $E^{Ising}_{dw}$
when $K\rightarrow\infty$: this is because it derives from a continuum
approximation, which fails to reproduce the discontinuity of an
Ising domain wall. However, when the Ising limit is pertinent, domains
are very large and probably unphysical. So, this limit is not really
relevant (at least at $T=0$).

An approximate analytical expression for $L$ can be found by noticing
that the dipolar interaction between domains can be evaluated $-$in the
Heisenberg model$-$ by ``removing" the domain wall~\cite{KP}: so, we have Ising
domains left, at distance $w$. By using Eq.~(\ref{Ic_2}) and the espression
given in Section~\ref{ss_dw} for the energy of a domain wall, we have the
followig formula for the energy per spin of a stripe domain configuration:
\be
E_{dom} = {\lambda w\over L} + {J\over Lw} - {4g\over L}\ln\left({L
\over 2w}\right)~,
\label{Ed_H}
\ee
which must be minimized not only with respect to $L$, but also $w$.
The calculation gives:
\bea
w &=& {16\pi L_d\over 2\left[ 1+\sqrt{1+4^3\pi^2 L_d(f-1)}\right] }\nonumber\\
L &=& 2w\exp({J/2gw})~.
\label{wL}
\eea
This expression for $L$ is valid also for the Ising model, where $w=1$
The form of $w$ proves the existence of a minimum value $f_m$, for which the 
square root vanishes and $L\simeq L_d$.
The value of $f_m$ found {\it via} the approximate expressions given above is 
not the correct one: $f_m=1-1/16L_d$, because Eq.~(\ref{Ed_H}) neglects the 
parameter $s$.

\section{Domain size and film thickness}
\label{LvsN}
\subsection{Kittel's theory and modifications}
The theories developped in the previous section apply to a single
monolayer. Anyway, before extending them to a (thin) film of $N$ planes,
let us report Kittel's old results~\cite{Kittel}, 
which are valid for a thick film,
whose thickness is larger than the domain size $(N>L)$, and a recent
modification~\cite{Zhu}, introduced to take into account the possible 
surface character of the anisotropy $K$.

\begin{figure}
\centerline{\psfig{%
bbllx=280bp,bblly=100bp,bburx=540bp,bbury=750bp,figure=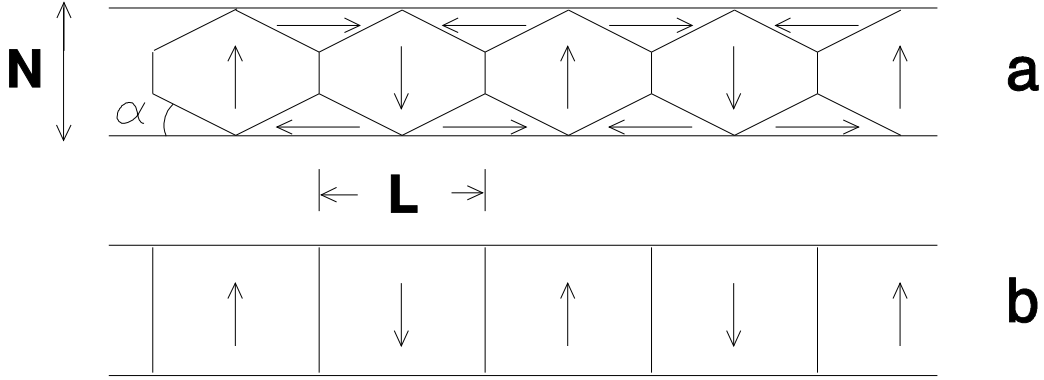,%
height=6cm,angle=90}}
\caption{Domain structures in a thin film: flux closure (a) and open flux
circuit (b).}
\label{fig_K}
\end{figure}

In Kittel's theory, the total energy $(F_{dom})$ of the magnetic slab is the
sum of three contributions: the magnetostatic one $(F_{dip})$, the
domain-wall energy $(F_{dw})$, and the anisotropy term $(F_{an})$.
He considered 
both the possibility of flux closure (Fig.~\ref{fig_K}a), and of an open flux
circuit (Fig.~\ref{fig_K}b), 
because he was interested in determing the lower energy
configuration. However, the dependence $L(N)$ is the same for the
two cases. If closure domains are present, the contribution of the
magnetostatic energy is neglected and the anisotropy term is
proportional to the volume of the domains where the magnetization lies in the
film plane. Finally, $F_{dw}$ depends on the extension of the domain
walls. If $\sigma_{w1}$ and $\sigma_{w2}$ are the wall energies per
unit area, respectively of the $90^\circ$ and $180^\circ$ walls,
the total energy $F_{dom}^{(a)}$ (per unit of surface) of the closure
domain structure, writes\footnote{The superscripts $^{(a)}$ and $^{(b)}$
refer, respectively to Fig.~2a and Fig.~2b.}:
\be
F_{dom}^{(a)}=F_{dw}^{(a)}+F_{an}^{(a)}\equiv
\left[ {2\sigma_{w1}\over\cos\alpha} +\sigma_{w2}\left(
{N\over L}-\tan\alpha\right)\right] + K{L\over 2}\tan\alpha~.
\label{F_K}
\ee
Kittel considered the limit $N\gg L$ and assumed $\sigma_{w1}=\sigma_{w2}$
and $\alpha=45^\circ$.

The minimization of $F_{dom}^{(a)}$ gives $L=\sqrt{2\sigma_{w2}N/K\tan\alpha}$
and $F_{dom}^{(a)}=(2\sigma_{w1}/\cos\alpha)-\sigma_{w2}\tan\alpha+
\sqrt{2K\tan\alpha\sigma_{w2}N}$. So, the domain size is
proportional to the square root of the film thickness.

Such a result is found also in absence of closure domains: in such case
$F_{dom}^{(b)}=F_{dw}^{(b)}+F_{dip}^{(b)}$, because $F_{an}^{(b)}=0$.
Nevertheless, the relevant functional dependences are unchanged:
$F_{dw}^{(b)}=\sigma_{w2}N/L$ and $F_{dip}^{(b)}\approx M_s^2 L$
($M_s$ being the magnetization of saturation),
i.e. it has the same $L$-dependence of $F_{an}^{(a)}$. Consequently,
the result $L\sim\sqrt{N}$ still holds.

Recently, the above calculation has been modified~\cite{Zhu} to take into
account the possibility of a surface anisotropy, in an ultrathin film.
In this limit, $N\ll L$, closure domains are supposed to make a small
angle $\alpha$ with respect to the surface of the film, and the total
anisotropy $K$ is written as the sum of a volume $(K_v)$
and a surface $(K_s)$ part:
$K=K_v+{2K_s/N}$. 
If $K_s(\approx 1$~erg/cm$^2$) prevails over $K_v(\approx 10^5$erg/cm$^3$),
the result by Kittel is
modified as follows\footnote{In Ref.~\cite{Zhu}, the trigonometrical
factor $\tan\alpha$ is missing.}:
\be
L=\sqrt{2\sigma_{w2} N\over K\tan\alpha}\simeq\sqrt{\sigma_{w2}\over
K_s\tan\alpha} N~.
\ee
So, for a surface anisotropy, Kittel's theory gives a {\it linear}
increase of the domain size with the thickness.
This result agreed with some experimental findings~\cite{AS} on thin epitaxial
Co/Au(111) films. Nevertheless, the calculation in Ref.~\cite{Zhu}
does not seem to be consistent.

In fact, it is noteworthy that below a critical thickness, Kittel's 
theory predicts a single-domain in-plane state. This result is obtained by
comparing the energy $F_\parallel=2K_s+K_vN$ of a single-domain
in-plane state with the energy of the closure domain state:
\be
F_{dom}^{(a)}={2\sigma_{w1}\over\cos\alpha} + 2\sqrt{K_s\sigma_{w2}\tan\alpha}
-\sigma_{w2}\tan\alpha~.
\ee
If we write $\sigma_{w1}\simeq\sigma_{w2}\simeq\sigma_w$ and we neglect the 
last two terms with respect to the first one (because $\tan\alpha\ll 1$), 
we find that 
the in-plane magnetized state is energetically favoured if\,%
\footnote{All the energies must be expressed in units where the
lattice constant $a=1$.} 
$N<2(\sigma_w-K_s)/K_v$: a condition which is generally fulfilled
in ultrathin films.

A more rigorous consideration is the following. Expression (\ref{F_K})
for the energy $F_{dom}^{(a)}$ 
of a closure domain structure should be minimized 
not only with
respect to $L$, but also to the angle $\alpha$.
If $\partial_\alpha F_{dom}^{(a)}$ is calculated, we obtain
\be
\partial_\alpha F_{dom}^{(a)} = {1\over\cos^2\alpha}\left[(KL/2 -\sigma_{w2})
-2\sigma_{w1}\sin\alpha\right]=0~.
\ee
However, it is easily seen that its solution: $\sin\alpha=(KL/2-\sigma_{w2})
/2\sigma_{w1}$ does not correspond to a local minimum: the Jacobi
matrix of the second derivatives has a positive and a negative eigenvalue, 
because $F_{dom}^{(a)}$ has a minimum with respect to $L$, 
but a maximum by varying $\alpha$!
Indeed, this problem exists also for the original Kittel's theory, where
$\alpha$ is arbitrarily put equal to $45^\circ$. A correct approach should
probably add an $\alpha$-dependent magnetostatic contribution 
$F_{dip}^{(a)}$.

\subsection{Ultrathin film limit: microscopic approach}
In the following, we will be concerned with the extension of the Ising
and Heisenberg models to the case $N>1$. 
Whilst the extension to $N>1$ of the domain wall energy is straightforward
for both models, the contribution of the dipolar interaction 
energy between different
planes is not trivial, and some further approximation is necessary, in
order to obtain tractable formulas. 

Let us start with the Heisenberg model, 
which is even simpler than the Ising one.
The domain wall size $w$ varies in the interval $\sqrt{J/K}<w<J/g$.
So, the hypothesis $w\gg N$ is reasonable, and in this limit the
integral $I(c)$ in Eq.~(\ref{Ic_2}) does not depend on $c$. 
Such integral appears $N$ times for the value $c=0$ and $N(N-1)/2$
times for $c\ne 0$, that is to say a total of $N(N+1)/2$ times.
Therefore, the expression (\ref{Ed_H}) for the
energy of the domain structure is modified as follows:
\be
{E_{dom}(N)\over N} \approx {\lambda w\over L} + {J\gamma(N)\over Lw}
-{2(N+1)g\over L}\ln\left({L\over 2w}\right)~.
\ee

The factor $\gamma(N)$, whose importance will be elucidated very shortly,
has been introduced to take into account a possible ``long-range"
character of the exchange interaction. If the range $r_{ex}$ of $\Hex$ is
much smaller than $N$, then $\gamma(N)=1$, but $\gamma(N)=
\gamma^*(N)\equiv (N+1)/2$ if $r_{ex}>N$.
Finally, the anisotropy term $\Han$ has been assumed to be of 
volume type.

The $N$-dependence of $E_{dom}$ can be reduced to the following
transformations: $J\rightarrow\gamma(N)J$ and $g\rightarrow \gamma^*(N)g$.
The domain wall size $w$ has no trivial transformation, but in the limit
$L\gg w$, $w\rightarrow\sqrt{\gamma(N)}w$ (see Eq.~\ref{wL}). So:
\be
L(N)\approx\sqrt{\gamma(N)} w_1 \hbox{e}^{{\sqrt{\gamma(N)}
\over \gamma^*(N)}{J\over 2gw_1}}~,
\ee
where $w_1$ refers to the monolayer $(N=1)$.
Since $\sqrt{\gamma(N)}/\gamma^*(N)$ 
is always a decreasing function of $N$, a strong
shrinkage of the domains is expected, even if $\Hex$ is ``long-range".

Now, let us consider the Ising model. When we deal with the dipolar term,
it is no more possible to suppose $w\gg N$, because now $w=1$. The integral
(\ref{Ic_2}) writes:
\be
I(c)\simeq 2 + (2/c)\arctan(1/c) + 2\ln(L/2\sqrt{1+c^2})
\approx 2\ln(L/2\sqrt{1+c^2})~,
\ee
if $L\gg c$. The total dipolar energy gain $E_{dip}$ is obtained by summing
up $I(c)$ over $c$, for all the couples of planes. Kaplan and
Gehring~\cite{KG} have used a continuum theory by resorting to the
magnetic potential: this corresponds to the limit $N\gg 1$. In this
case $I(c)\approx 2\ln(L/2c)$ and
\be
{E_{dom}\over N} \approx {2J\gamma(N)\over L} - 
{4Ng\over L}\ln\left({L\over N}\right)~.
\label{Ed_I}
\ee
The factor $N$ inside the logarithm does not modify the leading behaviour,
but in opposition to the Heisenberg model, it becomes relevant if
$\gamma(N)=\gamma^*(N)$. In fact, the minimization of (\ref{Ed_I}) gives
\be
L(N)\approx N\hbox{e}^{{\gamma(N)\over N}{J\over 2g}}~.
\ee
Now, if $\gamma(N)=\gamma^*(N)$, the exponential dependence on $N$
disappears\footnote{In the limit $N\gg 1$, the distinction between $N$ and
$(N+1)$ is not relevant.} and the polynomial dependence becomes relevant:
therefore, a linear increase of the domain size is predicted 
(in the Ising model!) if $r_{ex}>N$.

We can summarize the main results in the following way. In both models 
(Ising and Heisenberg models), $L$
has an exponential dependence on $(J/2gw)\approx(E_{dw}/g)$.
When the thickness is increased, in very rough terms 
$g\rightarrow \gamma^*(N)g$, but $E_{dw}$ has a
different expression in the two models: $E_{dw}\approx J\rightarrow\gamma(N)J$
in the Ising one, and $E_{dw}\approx\sqrt{J\lambda}\rightarrow\sqrt{\gamma(N)}
E_{dw}$ in the Heisenberg one. So, in the limit $\gamma=\gamma^*$, the
exponential dependence on $N$ disapperas in the Ising model, but it remains
in the other one. This dependence implies that $L$ decreases with $N$
in the Heisenberg model, whilst 
in the Ising model, if the exponential doesn't depend
on $N$, we have the pre-exponential dependence left: a
dependence which makes $L$ increase with $N$ and which becomes a linear
dependence in the continuum approximation.

We can ask how the result for the Heisenberg model is modified, if the
anisotropy is of surface-type: since the domain wall energy decreases with $N$,
narrower domains are expected! This result contradicts the findings on the
closure domain structures~\cite{Zhu}, based on the modified Kittel's
theory. The reason is the following: in this theory, the anisotropy 
contributes to the total energy through the closure domains,
where the magnetization lies in the film-plane, whilst the domain-wall
energy is tacitly supposed not to depend on it. 

Instead, in our theory no closure domain is present $-$as expected,
if the film is very thin$-$ but the domain wall energy is
proportional to the square root of the anisotropy.

The conclusion of the present section is that $-$as a general rule$-$ 
in a thin film domains
are expected to get narrower upon increasing the thickness. This behavior 
has been found experimentally by Speckmann~et~al.~\cite{Speckmann} in
ultrathin Co/Au(111) films (above $N=4$), by Bochi~et~al.~\cite{Bochi} in
Ni/Cu/Si(001) films, and by Gehanno and Samson~\cite{GS}
in ordered FePd thin films. So, a contradiction seems to exist between 
Ref.~\cite{AS} and Ref.~\cite{Speckmann}, both concerning
Co/Au films. This contradiction might be
solved guessing that the linear increase observed in Ref.~\cite{AS}
accords with the weak increase of Ref.~\cite{Speckmann}, oberved below
$N=4$. This behavior would correspond to the regime $N<r_{ex}$, which
gives $-$for an Ising model!$-$ an increasing $L(N)$. Above four monolayers,
we entry in the regime $N>r_{ex}$ (a narrow regime which is missed in
Ref.~\cite{AS}), which is characterized by a strong decreasing of $L(N)$.

Anyway, the previous picture is not totally satisfactory, because it
assumes the validity of the Ising model, or $-$in other words$-$ 
of domain walls
of atomic thickness, whilst walls extend over hundreds of lattice constants.
Furthermore, all the previous ``magnetic" models assume perfectly flat and
periodic arrays of spins, even if defects are known to play a major role
in determing domain structures. 
For example, Allenspach and Stampanoni~\cite{AS}
report an increase of the average lattice constant, upon increasing $N$.
This means that dislocations exist in the film, and also that the exchange
coupling constant is expected to decrease. Since $J$ strongly determines
the actual value of $L$, we can even suppose that the observed increase
of $L(N)$ is indeed due 
to the lowering of $J$, and therefore of the domain wall energy.

\section{Domain structures and reorientation transition}
\label{RPT}
\begin{figure}[t]
\centerline{\psfig{%
bbllx=100bp,bblly=100bp,bburx=560bp,bbury=750bp,figure=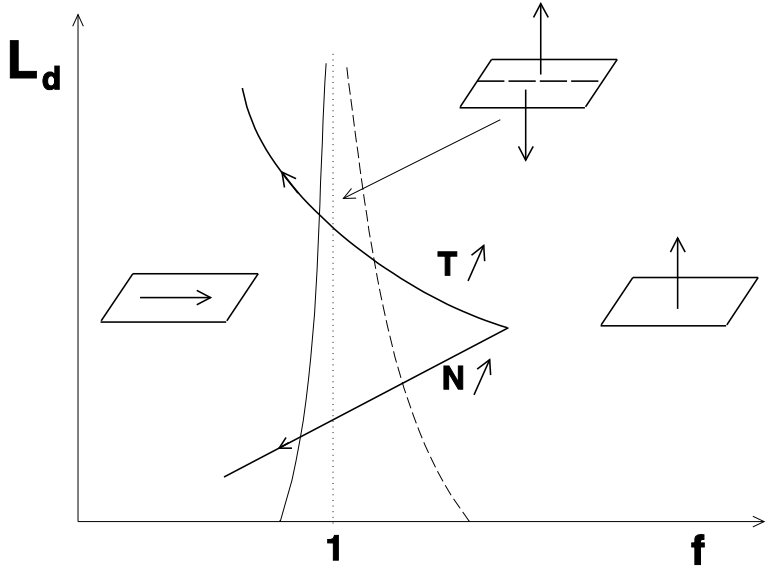,%
height=8cm,angle=90}}
\caption{``Phase diagram'' at $T=0$,
in the plane $(f,L_d)$. No units are shown on the
$f$ and $L_d$ axis, because the diagram has only a qualitative meaning.
The dotted line $f=1$ corresponds to the separation between the $\perp$
and $\parallel$ FM configurations. By taking into account domain
structures, domains rise in all the region on the right of the full thin line,
but they are observable only on the left of the dashed line.
So, starting from a thin film which is perpendicularly magnetized, we may
observe a reorientation phase transition upon increasing the temperature,
or upon increasing the thickness.
In a region around $f=1$, domains of observable size appear.}
\label{fig_P}
\end{figure}

In the part of the previous section which was dedicated to the
Heisenberg model, we shew the possible different ground states of an
ultrathin ferromagnetic film, subject to an easy-axis anisotropy and to the
dipolar interaction. In Fig.~\ref{fig_P}, we show the phase diagram at $T=0$,
in the space of the parameters $f=K/\Omega$, $L_{d}=J/\Omega$.
Upon decreasing the ratio $K/\Omega$, 
the system passes from a collinear $\perp$
state, to a collinear $\parallel$ state, across a domain structure.

As we have pointed out, the $\perp$ state is never the ground state,
in the thermodynamic limit. Anyway, from an experimental point of view, 
the question
is different. First of all, domains may be even larger than the sample size,
so that $-$in principle$-$ the $\perp$ state may be the true ground state
of a finite system. Furthermore, our estimates of the domain size were
valid at $T=0$, for a perfect sample.
Whatever the domain size is, the $\perp$ state $-$which is metastable$-$
is attained by the application of a magnetic field pulse: a typical procedure,
before a magnetization measurement. Therefore, an important quantity
is the relaxation time of the collinear state: if it is (much) longer than 
the measurement time, the $\perp$ state 
may be considered as the ``ground state", whereas
it will decay in a domain structure, in the opposite limit.

Experimentally, the phase diagram in the parameter space may be ``visited"
by varying the thickness and/or the temperature. In fact, the easy-plane
effect of the dipolar interaction increases with $N$, whilst the easy-axis 
anisotropy generally keeps constant, because of its surface character.
This means that the ``effective" ratio $K/g$ decreases, and indeed a
transition from the $\perp$ state to the $\parallel$ one is observed,
when the thickness is augmented. This transition is called reorientation
phase transition (RPT), and if a microscopic probe is used to measure the
magnetization (for example, the spin polarized scanning electron
microscopy), magnetic domains are observed in the ``transition" region
between the two collinear states~\cite{AB}.

The same phenomenon of reorientation may be observed upon increasing the
temperature, before the usual ferro-paramagnetic transition takes place.
In this case, the theoretical study of the transition requires to take
into account thermal fluctuations, which weaken the different coupling
parameters $(J,K,\Omega)$. The RPT is the result of a strongest decrease
of the uniaxial anisotropy, with respect to the dipolar coupling: so,
$K(T)$ may be smaller than $\Omega(T)$ $-$and therefore the $\parallel$ state
the preferred one$-$ even if $K(T=0)>\Omega(T=0)$.
The $T$-dependence may be obtained through a perturbation theory~%
\cite{EM,LG},
or through a renormalization group analysis~\cite{PP}.
We will not enter into details here, nor we will give the explicit
dependences on $T$; however, we want to discuss an important feature of
the RPT: the order of the transition.

Firstly, let us address the question in the parameter space.
If we consider the true ground state of the system, the answer is clear:
the transition is continuous, because the domain structure (see Fig.~1a), 
in proximity to
$f=1$, evolves towards the $\parallel$ state, passing through a shrinking
of the domain size and the {\it continuous} change of $s$ (the maximal value
of $|S_z|$ inside each domain) from $s=1$ to $s=0$. This picture clearly
shows the importance of introducing such variational parameter in the
study of domain structures.

If domain structures are neglected, as in almost the whole body of 
theoretical papers on
the RPT, the answer is more subtle, and $-$above all$-$ it depends on details
of the system. Limiting the investigation to collinear configurations
corresponds, in the $\perp$ phase, to study a metastable 
state\footnote{It should be 
stressed that in the (very) narrow interval $f_m<f<1$,
there is no collinear state which is locally stable.}.
Let us start with the simplest possibility: a single monolayer,
subject to the Hamiltonian ${\cal H}=\Hex +\Han +\Hdip$. We have shown that
$\Hdip$ has an effective easy-plane effect $(\Hdip^{an})$: as a 
consequence of this, 
the energy (per spin) of a collinear configuration writes:
\be
E(\theta)=-(K-\Omega)\cos^2\theta~,
\ee
where $\theta$ is the angle between the
magnetization and the $\hat z$ axis. 
It is manifest that $E(\theta)$ allows only two
possible minima, according to the sign of $(K-\Omega)$: $\theta=0$ and 
$\theta=\pi/2$. So, in the parameter space, the transition is obviously 
discontinuous, because no canted configuration $(0<\theta<\pi/2)$ is
allowed: so, it is not possible to change with continuity the value
of $\theta$ which minimizes $E(\theta)$, from $\theta=0$ ($\perp$ state)
to $\theta=\pi/2$ ($\parallel$ state).
Nevertheless, this result may be modified for two reasons:
(i)~the introduction of higher order anisotropies~\cite{ASB}, 
and (ii)~an increase
of the thickness~\cite{MU}.

In the former case, a term like $K_4\cos^4\theta$
is added to $E(\theta)$, and consequently $E'(\theta)=0$ allows solutions other
than $\theta=0,\pi/2$. This means that the minimum is expected to change
continuously from $\theta=0$ to $\theta=\pi/2$, upon varying the parameters.
In the latter case $(N>1)$, a possible canting of the magnetization is 
determined by the break of translational invariance in the perpendicular
direction, due to the surface character of the
easy-axis anisotropy~\cite{MU}, 
or even by the dipolar interaction itself, whose
easy-plane effect varies a little from plane to plane~\cite{PPR}.

From an experimental point of view, case (i) is much more important:
in fact, the extension of the region where canting is really observed
depends on $K_4/K$ in the first case, and on $\Omega/J$ 
$-$or $K/J$$-$ (two much smaller
quantities) in the second one. In other words, variations of the
effective anisotropy from layer to layer change the order of the
RPT (from discontinuous to continuous), but the transition from $\perp$
to $\parallel$ keeps very sharp.

Till now, we have analyzed the order of the RPT in the parameter space,
at $T=0$. What happens if the transition is induced 
by an increase in temperature?
We will consider the simplest case: a single monolayer in absence of 
fourth-order anisotropies. In fact, this is the only case which shows a first
order transition in the parameter space.

We have said that spin-wave interactions lead to a renormalization of the
coupling parameters: $(J,K,\Omega)\rightarrow(J(T),K(T),\Omega(T))$.
Anyway, if the {\it form} of the Hamiltonian keeps the same, the
energy of a collinear configuration: $E(\theta,T)=
-(K(T)-\Omega(T))\cos^2\theta$
cannot have other minima than $\theta=0,\pi/2$, and therefore no canted
configuration is possible. This is the result of several authors~%
\cite{PP,Teitelman,EM,Comment}, but it does not agree with recent
calculations by Chui~\cite{Chui}. This author claims that a single
renormalization group equation for the dipolar coupling is not sufficient:
under the flow of the renormalization group, the Hamiltonian would
acquire an additional term which changes the order of the transition,
as the introduction of a quartic anisotropy made the
transition a continuous one. Unfortunately,
the lack of details in Ref.~\cite{Chui}
prevents from drawing any conclusion.

\section{Domain structures at finite temperature}
\label{finiteT}
We have seen how the competition between $\Hex$ and $\Hdip$ gives rise
$-$at $T=0-$ to an ordered domain structure. This 
network of domain walls can be seen as a two dimensional solid, 
described by some positional and orientational order.
This solid will be seen to support long wavelength excitations
(wall meandering) as well as topological excitations (dislocations 
and stripe rotation domain walls).

In the Ising model, a single domain wall can be considered the
interface between regions of opposite magnetisation. In two dimensions,
the free energy $\sigma(T)$ of such an interface goes to zero at the critical 
temperature $T_c$ of the Ising model; nevertheless, it represents a one
dimensional interface, which is rough at any finite $T$.
In the present case, ``rough" means delocalized. This would imply that
the solid is always in a floating phase, with the walls which are depinned
from the substrate.

This is the {\it scenario} considered by Gehring and Keskin~\cite{GK},
who simply replace the domain wall energy, by its free energy.
Therefore $L\approx\exp{[\sigma(T)/g]}$ decreases with
$T$, till $L$ becomes of the same order of the domain wall size: from now
on, no regular structure exists and a disordered phase sets in.

Czech and Villain~\cite{CV} had pointed out that a floating phase exists only 
at temperatures higher than $T_R\approx g$. Indeed, neighbouring walls
interact both entropically and {\it via} the long-range dipolar coupling.
At sufficiently low $T$, entropic interactions can be neglected and a domain
wall consists of straight pieces of wall separated by kinks.
The typical distance between kinks is $\ell\approx e^{2J/T}$, where $(2J)$
is the energy of a kink. A ``coarse-grained" version of the system is
obtained by sampling the $m$-th wall at the points $y_p=p\ell'$, where
$\ell'<\ell$. The resulting dynamical variables are the displacements 
$\{ u_m^p\}$ of the sampled points, with respect to the $T=0$ positions. 
The value $\ell'$ is choosen sufficiently small so that 
$|u_m^p-u_m^{p+1}|\le 1$, but sufficiently large in order to suppose the
different pieces of wall as independent.
Even if kinks are absent $(|u_m^p-u_m^{p+1}|\equiv 0)$, the rigid
translation of a wall costs dipolar energy. The final expression for the
free energy writes~\cite{CV}:
\be
F=\sum_{m,p}\left[ J_\parallel (u_m^p-u_m^{p-1})^2 +
                   J_\perp     (u_m^p-u_{m-1}^p)^2 \right]~,
\label{F_CV}
\ee
which looks like an anisotropic Solid on Solid (SOS) model. 
The coupling  constant in the
wall direction $(J_\parallel\equiv T\ln(\ell/\ell'))$ 
is determined by the kink energy $(=T\ln\ell=2J)$ and by the 
entropy $S=\ln\ell'$ of each piece of wall.
On the other side, the coupling constant $J_\perp\equiv g\ell'/L^2$ 
is nothing
but $(\partial^2/\partial L^2)(-g\ln L)$ times $\ell'$.

It is easily seen that $J_\parallel\approx J_\perp\approx g$ when
$T\approx g$: this means that the roughening temperature of the solid
of domain walls is not $T_R=0$, but rather $T_R\approx g$.
At smaller temperatures, Czech and Villain suggest that $L$ changes
discontinuously (because of the discrete nature of the lattice), however
keeping itself a too large value to be experimentally observable.
At higher temperatures the network of domain walls represents a floating
solid.

We close this part by mentioning the mean-field result~\cite{CV}.
In this approximation, the free energy writes:
\be
F_{MF}=\sum_k (T-\tilde J(k))m_k m_{-k} + \tilde cT\sum_i m_i^4 + \dots
~,
\ee
where $\tilde c$ 
is a numerical factor and $\tilde J(k)$ is the Fourier transform
of the exchange and dipolar coupling between spins.

The value $k_m$ which minimizes the quadratic form is of order $(g/J)$, so 
that $L(T_C)\approx J/g$, whilst $T_C=\tilde J(k_m)\simeq J -O(g)$.
In other words, mean-field theory predicts relatively small domains
at the transition (of size of order of the dipolar length) and that 
$T_C$ is only slightly modified by dipolar interactions (at least in
the limit $g\ll J$).

The most detailed and complete study  on the different phases of a two
dimensional network of domain walls has been done by Kashuba and
Pokrovsky~\cite{KP} and by Abanov, Kalatsky, Pokrovsky and 
Saslow~\cite{Abanov} (hereafter called K2APS theory).
They consider a Heisenberg model: because of the possibility of a
Reorientation Phase Transition, the phase diagram is more complex.
Anyway, if we neglect the phases where the order
parameter is the magnetization, and we retain only the phase diagram of the
two dimensional solid of domain walls, we obtain a system 
which does not differ too much from the Ising case.
Indeed, in the limit $L\gg w$ it is possible to replace 
the Heisenberg domains,
separated by Bloch walls, with Ising domains, separated by empty walls.
Spin-wave fluctuations are taken into account through a renormalization
of the different coupling parameters. (This procedure is expected to be
correct if
$L$ is much larger than the length $L_R$ at which renormalizations stop.
Since this length is given by the relation $J/L_R^2\approx\lambda$,
the above condition equals $L\gg w$.)

Once the spins have no in-plane component, the ``volume" interaction
in $(1/r^3)$ between domains can be reduced $-$through the Gauss's theorem$-$
to an interaction in $(1/r)$ between domain walls. This allows to pass
from spin variables $(\vec S_i)$ to 
``displacement" variables $(u(x,y))$ of the
domain walls, and the energy of the system can be written as a functional
of the field $u(x,y)$.

In very general terms, we can say that the energy must contain three terms:
i)~A compression term, expressing the response of the system to a changement 
of $L$. This term is proportional to
$(\partial_x u)^2$. ii)~A bending term, expressing the energy due to a 
curvature $(\partial^2_y u)$ of the domain wall. iii)~A stripe orientation
term, expressing the existence of preferred directions for domain walls.
This term, proportional to $(\partial_y u)^2$, has an extreme importance
because its presence drastically alters the behaviour of the system.

By taking into account all the above mentioned terms, we obtain%
~\cite{KP,Abanov}:
\be
E_S=\int\int dxdy\left[{K_S\over 2}\left({\partial u\over\partial x}
\right)^2 + {\mu\over 2}\left({\partial^2 u\over\partial y^2}\right)^2
+ {\nu\over 2}\left({\partial u\over\partial y}\right)^2\right]~.
\label{E_K2APS}
\ee
The first term 	is the ``continuum" version of the term proportional to
$J_\perp$ in the SOS Hamiltonian by Czech and Villain~\cite{CV}.
In fact, according to K2APS theory, $K_S\approx \Omega/L$, 
which becomes, after the
transformation $x\rightarrow Lx$, $K_S\approx \Omega/L^2$. 
This expression equals
$J_\perp$, once we have put $\ell'=1$. 

We could therefore be induced to conclude that the third term corresponds
to the one proportional to $J_\parallel$ in the same SOS Hamiltonian.
Indeed, the question is more subtle, as suggested by the fact that $\nu$
is not simply proportional to the domain wall energy.
In fact, since $\theta\equiv\partial_y u$ is the local orientation
of the domain wall, in absence of anisotropic terms the energy $E_S$ can
not explicitely depends on it.

The microscopic nature of $\nu$ is the following: once we have fixed all
the parameters, the domain wall energy per unit length depends on the
orientation of the wall itself. If we consider an Ising model on a square
lattice, a domain wall in the direction (0,1) $-$or (1,0)$-$ is less
expensive that the same domain wall in the (1,1) direction:
$E_{dw}(0,1)=2J/a$ and $E_{dw}(1,1)=2J/(a/\sqrt{2})=\sqrt{2}E_{dw}(0,1)$,
where we have explicitly written the lattice constant $a$.
So, on a square lattice we will have a stripe orientation energy with
periodicity $(\pi/2)$. Its expansion in the vicinity of a minimum gives rise
to the third term of $E_S$. 

The correct expression of $\nu$ for the Heisenberg model is given in
\cite{KP}: $\nu=\nu_{ex}={1\over L}\left({J\over w^3}\right)$, where the factor
$(1/L)$ simply translates the continuum description of walls at
distance $L$.
We want to point out that $\nu$ may have also a different origin:
an in-plane anisotropy, for example. It is easily seen that in this
case $\nu=\nu_{an}\approx K_\parallel w/L$, where $K_\parallel$ is the
single-ion in-plane anisotropy. So, $\nu_{an}>\nu_{ex}$ if 
$(K_\parallel/\lambda)>(\lambda/J)$, a condition which may be fulfilled
in some systems. Furthermore, the periodicity of $\nu_{an}$ may be different
from that of $\nu_{ex}$ (see the end of this section).

Finally, the bending energy term (proportional to $\mu$) $-$which was
neglected in~\cite{CV}$-$ is proportional to the square of the curvature
$(\partial_y^2 u)$, because of the required invariance 
$u\rightarrow -u$.

Once we have explained the origin of the energy $E_S$, let us discuss
the $T$-dependent behaviour of our system. Two quantities are of interest:
the mean square fluctuations of the position $\langle u^2\rangle$, and the
mean square fluctuations of the angular deviations $\langle (\partial_y u)^2
\rangle$. The first ones are given by:
\be
\langle u^2\rangle = T\int\int {dp_x dp_y \over K_S p_x^2 +\nu p_y^2
+\mu p_y^4} = {2T\over\sqrt{K_S}}\int {dp_y\over\sqrt{\nu p_y^2 +\mu
p_y^4}} \arctan\sqrt{K_S\over\nu p_y^2 +\mu p_y^4}~.
\ee
Since an ultraviolet cutoff is naturally supplied by the discrete nature  
of the system $(|p_y|<1)$, we are mainly concerned with a possible infrared 
divergence. When $p_y\rightarrow 0$, the integral has always a divergence:
a logarithmic one if $\nu\ne 0$ and a power-like one if $\nu =0$. The
difference is relevant, because the calculation of the correlation function
$\langle u(0)u(\vec r)\rangle$ shows a power-like decay, and therefore the
existence of a quasi long-range order (QLRO), solely in the first case:
in absence of a stripe orientation energy, domains are completely
delocalised and the derivative $(\partial_x u)$ has even no meaning.
The calculation of $\langle(\partial_y u)^2\rangle$ simply needs to add a 
factor $(p_y^2)$ at the numerator of the previous integral: because of
this factor, the angular deviations are always finite, even if $\nu=0$.

By summarizing: in the ``real" case $\nu\ne 0$,
there is an orientational long-range order (LRO) and a positional
QLRO. A magnetic domain phase with these properties has been indeed 
observed by Allenspach and Bischof~\cite{AB} in an ultrathin film
of Fe/Cu(100), with stripe domains oriented along a high symmetry
direction in the plane. 

The absence of LRO for the position of domain walls apparently
clashes with the results by Czech and Villain. Anyway, no real
contradiction exists:
at the extremely low temperatures considered in~\cite{CV} $(T<\Omega)$,
excitations in the Ising model are given by kinks, and a discrete 
description of domain walls
is necessary. In this case, we obtain the SOS Hamiltonian 
(\ref{F_CV}), to which corresponds a finite depinning temperature.
Conversely, the picture given by K2APS theory is a continuous one, valid
at not too low temperatures $(T>\Omega)$, where even the Czech and Villain's
theory foresees a floating phase.

Let us come back to Eq.~(\ref{E_K2APS}):
at sufficiently large lengthscales, the higher order term $(\partial_y^2
u)^2$ is negligible, and $E_S$ keeps an anisotropic quadratic form.
This energy is well known~\cite{Jancovici} to correspond to an
algebraical decay of the fluctuations of the order parameter.
The corresponding solid is called a smectic crystal by K2APS:
the QLRO is destroyed by the proliferation of pairs of dislocations,%
\footnote{A pair of dislocations of opposite Burger vectors rises
from the ``collapse" of a piece of two neighbouring walls.}
and the corresponding critical temperature writes according to
the Kosterlitz-Thouless expression~\cite{KT} ($P$ stands for
``positional"): $T_P={1\over 8\pi}K'b^2={\sqrt{K_S\nu}\over 2\pi}L^2$,
where $K'=\sqrt{K_S\nu}$ is the (renormalized) effective elastic 
constant and $b=2L$ is the
Burger vector corresponding to the dislocation.

Above $T_P$ the compression coefficient vanishes and domain walls are
delocalized. What happens to the orientational order parameter
(hereafter explicitely written through the angle $\theta$)?
The bending energy rewrites as ${\mu\over 2}(\partial_{y'}\theta)^2$,
whilst for the orientational stripe energy the complete expression
will be used: ${\nu\over 16}(1-\cos 4\theta)$. We must also add a term of the
form ${\kappa\over 2}(\partial_{x'}\theta)^2$, which would correspond to
${\kappa\over 2}(\partial^2_{xy}u)^2$ in the ``old" coordinates and which
was irrelevant to determine the large scale properties of $E_S$.
Finally, we will have~\cite{Abanov}:
\be
E_N=\int\int dx' dy'\left[ {\kappa\over 2}\left({\partial\theta\over
\partial x'}\right)^2 + {\mu\over 2}\left({\partial\theta\over
\partial y'}\right)^2 + {\nu\over 16}(1-\cos 4\theta)\right]~,
\ee
where $(x',y')$ are local coordinates.

An important point is the sign of $\kappa$, which may be positive or negative:
its mean-field calculation gives $\kappa_{mf}\approx -\Omega L$, but both
thermal fluctuations and corrections due to free dislocations 
give~\cite{Abanov} a positive contribution to $\kappa$.
An orientational ordered phase will exist at $T>T_P$, only if $\kappa>0$.
In this case,
because of the anisotropic term proportional to $\nu$ $-$which assures an
orientational LRO$-$ the order parameter $\theta$ has a discrete Z$_2$
symmetry:
domain walls, in a square lattice, can be directed either along
a given direction, or along the direction perpendicular to it.
For this reason, this phase has been called Ising nematic phase. 
At temperatures slightly larger than $T_P$, the two directions are not
equally probable and the Z$_2$ symmetry is broken.
Two contiguous regions where domains are directed along the two
different ``easy-axes", are separated by a so-called 
stripe rotation domain wall, 
studied in detail by Abanov et al.~\cite{Abanov}.
At a temperature $T_0>T_P$ we therefore expect an Ising-type transition
towards a liquid phase. Since there will be two (mutually perpendicular)
directions which are more likely, this phase is called a 
``tetragonal liquid". In this phase, square domains should be still
observable, with a magnetization inside each domain which is
perpendicular to the surface. This phase may evolve towards a planar phase, 
{\it via} a Reorientation Phase Transition, or directly towards a
paramagnetic phase. 

For a negative $\kappa$, the Ising nematic phase does
not exist and there is~\cite{Abanov} a first-order phase transition between
the (smectic) crystal and the (tetragonal) liquid.

Till now, we have supposed that the anisotropic stripe energy
has a $(\pi/2)$-periodicity and therefore domains orient
preferentially along two perpendicular directions.
We can suppose that in certain cases the in-plane symmetry is
reduced, for example because of an uniaxial distorsion, which
determines a quadratic spin anisotropy (which looks like
$\Han^\parallel=-{K_\parallel\over 2}\left.\int d\vec x S_y^2\right)$ or a 
modification of $\Hex$ such that the exchange coupling $J$ differs
in the two spatial directions $x$ and $y$.
In both cases, the existence and the nature of the smectic-nematic
transition at $T=T_P$ are not changed, because for $T<T_P$ only the
expansion at small $\theta$ of the stripe energy is relevant, not its
global periodicity. Conversely, for $T>T_P$ we are interested in the
orientational order, and for a $\pi$-periodicity the resulting
scenario is completely different. In fact, in this case there is
only one preferred direction for domain walls and no symmetry breaking
transition may take place at $T=T_O$.
The system behaves as a nematic in an electric field. The orientational order
is not destroyed, till a ``magnetic" transition towards a planar phase,
or a paramagnetic phase, occurs.

It is interesting to compare the given picture with the results of a
Monte~Carlo simulation on an Ising square model, due to 
Booth et al.~\cite{Booth}. The authors study a $64\times 64$ spins sample
with a ratio $J/\Omega\simeq 9$, such that $L\simeq 8$ at $T=0$.
They single out three different phases, through the existence of two
maxima in the specific heat $C$. By the comparison with Monte~Carlo
configurations at different temperatures, they conclude that there is
(I)~A low-$T$ phase with an orientational order of domain walls;
(II)~A middle-$T$ phase with no orientational order, but still well
defined domains; (III)~A high-$T$ phase, which is fully disordered.

Authors suggest that (I) resembles the smectic crystal phase, and (II)
the tetragonal liquid one. Since their model can not undergo a RPT,
(III) is the paramagnetic phase. Authors also add that such parallel
suffers a weak point: the transition between (I) and (II) seems to be
continuous, whilst the transition between the smectic crystal and the
tetragonal liquid would be first-order, or there would be a further
phase (the Ising nematic one) in between.

In this respect, some comments are in order. Authors~\cite{Booth}
study only the orientational order, and no details are given about
the positional one. Besides, domain size is rather small and 
dislocations are not visible in the figures reported in \cite{Booth}.
However, the most important point is the nature of the transition
between the smectic crystal and the Ising nematic phase. This transition
corresponds to the loss of the positional QLRO and has the nature
of a Berezinsky-Kosterlitz-Thouless transition. Such transition,
linked to the unbounding of pairs of dislocations, is not
``visible" through the specific heat $C$, because it corresponds
to an unobservable essential singularity  and $C$ has no
maximum for $T=T_P$.

We argue that the observed phase transition between (I) and (II) should
correspond to the transition between the Ising nematic phase and the
tetragonal liquid.

\section{Conclusions}
In this paper we have reviewed theoretical results on
magnetic domains in ultrathin films. Two-dimensional domain structures have 
a more fundamental meaning than the corresponding ones in bulk systems:
in fact, in three dimensions domains strongly depend on the shape of
the sample, and in a given sample are generally present for any
direction of the magnetization\footnote{With the exception of a toroidal
sample.}. Conversely, in two dimensions shape effects are negligible in the
thermodynamic limit, but domains are absent if the magnetization lies in the
film plane. Furthermore, the interplay between
uniaxial anisotropy and shape anisotropy possibly determines a
reorientation transition: domains are expected to appear just before
such transition takes place (upon increasing the thickness and/or the 
temperature).

The main limit of the present article is to have neglected the possible
application of a magnetic field. Because of the ``reduced" quantity of
matter, in a thin film the energy gain of a domain structure with respect to 
a collinear state is extremely low (see Fig.~1b). As a consequence of this,
a lower magnetic field is necessary to obtain a single domain state.
%a very low magnetic field $H$ is sufficient to obtain a single domain state:
%$H\approx 1\div 10$~Oe instead of $H\approx 10^3$~Oe, as in a bulk system.

A theoretical comprehension of the $T$-dependence of the network of domain
walls has received an important impulsion with the K2APS theory, but the 
possibility to discover in a real system the proposed sequence of phase
transitions is limited by the smallness of the region of temperatures
where domains are observable.
Furthermore, comparison of theoretical predictions with experimental
data is made more difficult in ultrathin films by the increased importance
of defects (dislocations and vacancies) and by the roughness of the
interfaces.

Finally, simulation data start to be available, but they often concern
Ising systems, and domain sizes are generally far from being
realistic.


\begin{thebibliography}{99}
\bibitem{Weiss}
P. Weiss, J. de Phys. {\bf 6}, 661 (1907). 
\bibitem{Bitter}
F. Bitter, Phys. Rev. {\bf 38}, 1903 (1931).
\bibitem{APA}
See, for example, the two special issues on {\it Magnetism in ultrathin
films} (D.~Pescia~Ed.): Appl.~Phys.~A {\bf 49}(5-6) (1989).
\bibitem{LKittel}
C. Kittel, {\it Introduction to solid state physics}, J.~Wiley 
(New~York, 1996).
\bibitem{CV}
R. Czech and J. Villain, J. Phys.: Condens. Matter {\bf 1}, 619 (1989).
\bibitem{KG}
B. Kaplan and G.A. Gehring, J. Magn. Magn. Mater. {\bf 128}, 111 (1993).
\bibitem{MacIsaac}
A.B. MacIsaac, J.P. Whitehead, M.C. Robinson and K. De'Bell, 
Phys. Rev. B {\bf 51}, 16033 (1995).
\bibitem{Maleev}
S.V. Maleev, Sov. Phys. JETP {\bf 37}, 1240 (1976)
[Zh. Eksp. Teor. Fiz. {\bf 70}, 2374 (1976)].
\bibitem{Bruno}
P. Bruno, Phys. Rev. B {\bf 43}, 6015 (1991).
\bibitem{PPR}
P. Politi, M.G. Pini and A. Rettori, Phys. Rev. B {\bf 46}, 8312 (1992). 
\bibitem{YG}
Y. Yafet and E.M. Gyorgy, Phys. Rev. B {\bf 38}, 9145 (1988).
\bibitem{Tesi}
P. Politi, Ph. D. Thesis (Florence, 1994). Unpublished.
\bibitem{Libro}
P. Politi, C.H. Back, M.G. Pini, A. Rettori and D. Pescia,
Contribution to {\it Fundamental aspects of thin film magnetism},
Eds. D. Pescia and A. Rettori, World Scientific (Singapore). To be published.
\bibitem{KP}
A.B. Kashuba and V.L. Pokrovsky, Phys. Rev. Lett. {\bf 70}, 3155 (1993);
Phys. Rev. B {\bf 48}, 10335 (1993).
\bibitem{Kittel}
C. Kittel, Phys. Rev. {\bf 70}, 965 (1946).
\bibitem{Zhu}
Zhu-Pei Shi, J. Phys.: Condens. Matter {\bf 4}, L191 (1992).
\bibitem{AS}
R. Allenspach and M. Stampanoni, Mat. Res. Soc. Symp. Proc. 
{\bf 231}, 17 (1992).
\bibitem{Speckmann}
M. Speckmann, H.P. Oepen and H. Ibach, Phys. Rev. Lett. {\bf 75}, 2035 (1995).
\bibitem{Bochi}
G. Bochi et al., Phys. Rev. Lett. {\bf 75}, 1839 (1995).
\bibitem{GS}
V. Gehanno, Y. Samson, A. Marty, B. Gilles and A. Chamberod,
J. Magn. Magn. Mater. {\bf 169} (1997).
\bibitem{AB}
R. Allenspach and A. Bishof, Phys. Rev. Lett. {\bf 69}, 3385 (1992).
\bibitem{EM}
R.P. Erickson and D.L. Mills, Phys. Rev. B {\bf 46}, 861 (1992).
\bibitem{LG}
A.P. Levanyuk and N. Garcia, J.Phys.: Condens. Matter {\bf 4}, 10277 (1992).
\bibitem{PP}
D. Pescia and V.L. Pokrovsky, Phys. Rev. Lett. {\bf 65}, 2599 (1990).
\bibitem{ASB}
R. Allenspach, M. Stampanoni and A. Bishof, Phys. Rev. Lett. {\bf 65},
3344 (1990).
\bibitem{MU}
A. Moschel and K.D. Usadel, Phys. Rev. B {\bf 51}, 16111 (1995).
\bibitem{Teitelman}
M.G. Tetel'man, Sov. Phys. JETP {\bf 71}, 558 (1990)
[Zh. Eksp. Teor. Fiz. {\bf 98}, 1003 (1990)].
\bibitem{Comment}
P. Politi, A. Rettori and M.G. Pini, Phys. Rev. Lett. {\bf 70}, 1183 (1993). 
\bibitem{Chui}
S.T. Chui, Phys. Rev. Lett. {\bf 74}, 3896 (1995).
\bibitem{GK}
G.A. Gehring and M. Keskin, J. Phys.: Condens. Matter {\bf 5}, L581 (1993).
\bibitem{Abanov}
Ar. Abanov, V. Kalatsky, V.L. Pokrovsky and W.M. Saslow,
Phys. Rev. B {\bf 51}, 1023 (1995).
\bibitem{Jancovici}
B. Jancovici, Phys. Rev. Lett. {\bf 19}, 20 (1967).
\bibitem{KT}
J.M. Kosterlitz and D.J. Thouless, J. Phys. C: Solid State Phys. 
{\bf 6}, 1181 (1973).
\bibitem{Booth}
I. Booth, A.B. MacIsaac, J.P. Whitehead and K.De'Bell, 
Phys. Rev. Lett. {\bf 75}, 950 (1995).
\end{thebibliography}
\end{document}